\documentclass[a4paper,11pt]{article}
\usepackage{jinstpub} % for details on the use of the package, please
\pdfoutput=1 % if your are submitting a pdflatex (i.e. if you have
\title{Development of Low Radioactive Molecular Sieves for Ultra-Low Background Particle Physics Experiment}

\author[a,d,1]{H.~Ogawa,\note{Corresponding author.}}
\author[b,d]{K.~Abe,}
\author[c]{M.~Matsukura}
\author[c,e]{and H.~Mimura}

\affiliation[a]{CST Nihon University, Surugadai, Kanda, Chiyoda-ku, Tokyo, 180-0011, Japan}
\affiliation[b]{Kamioka Observatory, Institute for Cosmic Ray Research, the University of Tokyo, Higashi-Mozumi, Kamioka, Hida, Gifu, 506-1205, Japan}
\affiliation[c]{UNION SHOWA K.K., A Joint Venture of UOP and SHOWA DENKO, Konan Minato-ku, Tokyo, 108-0075, Japan}
\affiliation[d]{Kavli Institute for the Physics and Mathematics of the Universe (WPI), the University of Tokyo, Kashiwa, Chiba, 277-8582, Japan}
\affiliation[e]{Professor Emeritus, Tohoku University}

\emailAdd{ogawa.hiroshi@nihon-u.ac.jp}

\abstract{This study develops low radioactive molecular sieves (MS) for ultra-low background particle physics experiment. The manufactured MS is of type 4A. $^{226}$Ra and $^{232}$Th have concentrations of about 57 and 198 mBq/kg, respectively, measured with high-purity germanium (HPGe) detector. The achieved reduction of the radioactivity with respect to the commercial one is about 99 and 97 percent for $^{226}$Ra and $^{232}$Th, respectively. Furthermore, the radon emanation rate from MS filter is about 2.1 mBq, which is measured by utilizing pin-photo type radon detector. This developed MS is expected to remove the impurity from the noble gas, in which low radioactivity is needed such as dark matter search experiment. }


\begin{document}
\maketitle
\flushbottom
\renewcommand\thefootnote{\alph{footnote}}
\section{Introduction}
\label{sec:intro}
Liquid xenon (LXe) has recently raised attention due to its application as noble gas detector media for ultra-low background particle physics, such as direct dark matter search experiments \cite{Aprile}. The scintillation light of LXe is as large as that of NaI. 
The wavelength of scintillation light is about 175 nm. Xenon purification is an important task in LXe detectors because the scintillation light is absorbed by impurities like, for example, oxygen and water. 

A major agent for impurity absorption is Molecular Sieve (MS), also called Zeolite, which is well known and commercially distributed. There are several types of MS such as 3A, 4A, 5A and 13X depending on the pore size. The uniform pore size is enough for the selective absorption of impurities. MEG experiments performed xenon purification by applying MS filtering and liquid circulation \cite{MEG}. MS can be used for purification of not only xenon but also other noble gases, such as liquid argon, which is also the detector media for dark matter search experiments. 
Besides, MS adsorbs radioactive impurities, for example radon \cite{CYG}, from the rare gases. Consequently, MS is useful in the purification of noble gases for dark matter search experiments.

For low background experiments, commercial MS includes a large amount of radioactive isotope, and especially beta and gamma rays originating from the decay of radon are the main background candidates for dark matter search experiments. Radon is originated as gas from the MS and then it diffuses into a detector. In order to optimally use a MS for purification in dark matter search experiments, the radioactive isotope in the MS has to be reduced.
 
This study develops low radioactive MS, which can be used for dark matter search experiments \footnote{This study is a joint research with UNION SHOWA K.K. \cite{Union}}. The radioactivity of the materials for MS and those developed is measured using HPGe and pin-photo type radon detectors.  The rest of this paper is organized as follows. Section~\ref{sec:target} demonstrates the target radioactivity of MS. Section~\ref{sec:develop} reports the actual development of 4A MS. Section~\ref{sec:meas} presents the achievement of low activity MS. Finally, Section~\ref{sec:conc} concludes the paper.  
  
\section{The Target of Radioactivity for MS Development}
\label{sec:target}
The typical target of LXe impurity (water, oxygen and others) is required to be $<$1 ppb. In this study, the performance of the developed MS is targeted to have the same level of adsorption capacity as that of commercial MS. Then, the amount of water absorption needed in MS more than 20 g for a 100 g MS. 

As radon source, $^{222}$Rn (with a half-life of 3.82 days) of uranium series and $^{220}$Rn (with a half-life of 55.6 s) of thorium series are considered.
In the case of dark matter search experiment, the activity of $^{222}$Rn should be less than 1 mBq/ton in a noble gas detector with the background level ${\rm 1 \times 10^{-4} \, kg^{-1} keV^{-1} day^{-1}}$ for $<$ 100 keV, which covers the energy region for dark matter search. Now we assume that amount of MS in purification filter is about 100 g. The radioactivity in MS requires 10 mBq/kg for $^{226}$Ra which is parent isotope of $^{222}$Rn. $^{220}$Rn has a very short half-life when compared with $^{222}$Rn. It is possible to reduce the decay of $^{220}$Rn in the detector space by, for example, installing the buffer space between detector and filter. Thus the target value of the activity can be relaxed. In this study, a target value of 100 mBq/kg or less in MS (with a margin, $^{222}$Rn target value $\times$ 10) was set for the concentration of $^{220}$Rn and $^{232}$Th. 

The radioactivity of commercial MS was measured using a HPGe detector. The radioactivities of $^{226}$Ra and $^{232}$Th in commercial MS is about 5.5 and 7.0 Bq/kg, respectively. Large amount of radon must be released from MS due to huge amount of radioactivity. It is necessary to develop a MS that reduces $^{226}$Ra to about 1/500 or less from a commercial MS. Commercial MS contains a binder made of natural clay to make the MS shape. This binder contains a lot of radioactivity. MS without binder still includes about 1.2 Bq/kg concentration of $^{226}$Ra. We select both the low radio-isotope (RI) material for MS and substitution of the binder.

\section{Material Selection and Development of MS}
\label{sec:develop}
The basic material for MS4A is NaOH, Al(OH)$_{3}$ and silica component. The selected NaOH was developed by Wako Junyaku Co. Ltd. and is a special grade reagent.  Al(OH)$_{3}$, also called BHP39, was developed by Nihon Keikinzoku Co. Ltd. and it is a further refinement of the imported Al(OH)$_{3}$. Meanwhile, we selected the Snowtechs ST-30 developed by Nissan Kagaku Co. Ltd., and about 30$\%$ silica component is included.  Silica component was also used as a solidifying agent instead of clay component. 
Table~\ref{table:selRI} presents the selected materials and their radioactivities for $^{226}$Ra and $^{232}$Th measured using the HPGe detector.  
The measured materials were transferred into a polyethylene (PE) vessel and the grove box was filled with pure air to guarantee a low radon level by avoiding contamination from the radon in the room air. Furthermore, vessels are inserted into the ethylene-vinyl alcohol copolymer (EVOH) bug. We prepared the empty PE vessel to evaluate the background in HPGe detector measurement. The concentration of $^{226}$Ra takes the value of the upper limit of NaOH and Al(OH)$_{3}$. We checked the various kinds of samples for Al(OH)$_{3}$. BHP39 is a lower concentration of $^{226}$Ra. The significant $^{226}$Ra and $^{232}$Th sources still remain in the silica component. Another candidate of silica component, which is reagent Na$_{2}$SiO$_{3}$ has 10 times the concentration of $^{226}$Ra compared with Snowtechs ST-30. This was selected. 
The water is required for the MS development. The ion exchange pure water was supplied from Organo Corporation. The concentration of uranium is less than 2 ppt measured using  Inductively Coupled Plasma Mass Spectrometry (ICP-MS) supplied by Toshiba Nanoanalysis Co.Ltd.

\begin{table*}[htbp]
\label{table:selRI}
\begin{center}
\caption{The selected materials and their radioactivities measured using HPGe detector.}
\begin{tabular}{lccl}
    \hline \hline
    material&$^{226}$Ra[mBq/kg]&$^{232}$Th[mBq/kg]&Company/Commercial name\\
    \hline  
   NaOH&$<$12.2&$<$8.14&WAKO/NaOH for precise analysis\\
   Al(OH)3&$<$9.1&$<$4.26&Nihon Keikinzoku / BHP39 \\
   Silica component&19.6$\pm$0.3&93.4$\pm$4.3&Nissan Kagaku/Snowtechs ST-30\\
    \hline \hline
\end{tabular}
\end{center}
\end{table*}

The new MS was developed at Nihon University with the supervision of UNION SHOWA K.K.. All the devises used for the development were washed with pure water to keep them clean and avoid contamination by outside impurities. The four steps for MS production were as follows. First we produced the aluminate configured from NaOH, Al(OH)$_{3}$ and pure water. The component of the developed aluminate was checked by specific gravity and titration measurement. Second, the crystal formation process of MS4A by mixing of aluminate, silica component and pure water. Third, the drying process for making MS powder. Fourth, the formation of shape by utilizing a solidifying agent with silica component and pure water using electric furnace. In these processes, the temperature and the time of sample stirring, drying, mixture etc., were controlled based on the recipes provided by UNION SHOWA K.K.. The development processes was performed three times and the developed MS was separated for each developed process. The total amount of developed MS was about 450 g.  

The ratio of crystallization and water adsorption capacity were measured for the developed MS4A powder by UNION SHOWA K.K., as presented  in Table~\ref{table:UNION}. 
In sample 2, the crystal formation process was different when compared with samples 1 and 3. Based on this, water adsorption capacity was reduced too but it was still enough. The performance of the developed MS was comparable to that of a commercial MS.

Sample 1 was solidified after the performance check and HPGe measurement in section~\ref{sec:meas}. We mixed silica component and pure water with the developed MS powder. The mixture sintered in an electric furnace, and the solidified sample was crushed. A sample of 120 g was collected.

\begin{table*}[htbp]
\label{table:UNION}
\begin{center}
\caption{The ratios of crystallization and water adsorption capacity for developed MS4A powders. These results were measured by UNION K.K..}
\begin{tabular}{lcccc}
    \hline \hline
    sample name&sample 1& sample 2 & sample 3 &commercial MS\\
    \hline
    water adsorption capacity [g/100g]&23.2&18.6&23.3&$>$ 20\\
    the ratio of crystallization [$\%$] & 69.4&53.7&70.0&$>$ 80\\
    \hline \hline
\end{tabular}
\end{center}
\end{table*}

\section{The Radioactivity Measurement of the Developed MS}
\label{sec:meas}
The radioactivities of the developed MS4A were measured by applying two methods. One is HPGe detector measurement, whereas the other is pin-photo type radon detector measurement. These devices are owned by Institute for Cosmic Ray Research, University of Tokyo and Kavli Institute for the Physics and Mathematics of the Universe, University of Tokyo, respectively.

For the HPGe measurement, the sample treatment followed the same method as in Section~\ref{sec:develop}. Since sample 3 was utilized to test the development of the other type MS, only samples 1 and 2 were measured. Additionally, sample 1 was measured again after it solidified and was referred to as sample 1b. Table~\ref{table:devMS4A} presents the results of the RI measurement. The results of samples 1 and 2 agree with the sum of the results of RI measurement of the material (Table~\ref{table:selRI}).  Moreover, sample 1b has a larger activity. The increase in the  $^{226}$Ra and $^{232}$Th in sample 1b can be explained by the content in silica component applied for solidification.

\begin{figure}[htbp]
  \begin{center}
    \includegraphics[keepaspectratio=true,height=45mm]{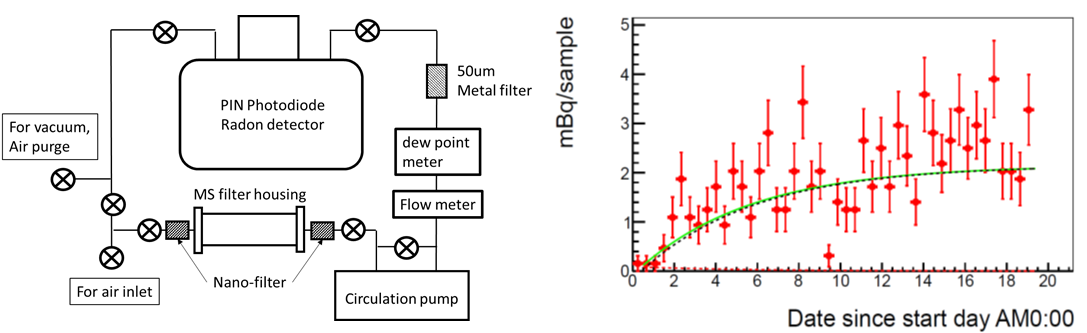}
  \end{center}
  \caption{Left: The setup of $^{222}$Rn emanation measurement. Right: Time profile of $^{222}$Rn emanation measurement for sample 1b. Red dots is measured emanation rate. Green, black dash and red dash lines indicate the fitting result of $R$, first term and second term in Equation~\ref{eq:fitting}, respectively.}
  \label{fig:ema}
\end{figure}

The radon emanation was measured for the MS filter after HPGe measurement. We placed 120 g of sample 1b in the filter housing, which comprises ConFlat (CF) nipples, metal particle filters and valves. The solidified MS was baked at about 200 $^{\circ}$C for 1 h in the filter housing to release water inside the MS. 
The left picture on Figure~\ref{fig:ema} shows the setup of $^{222}$Rn emanation measurement. Radon detector and filter housing are connected by Stainless Used Steel (SUS) pipes, valves, dew point meter, flow meter and air-tight sealing compressor. 

Radon detector is configured by applying pin-photo dynode with electronic field to capture the daughters of $^{222}$Rn ( $^{218}$Po and $^{214}$Po) \cite{RnDet}. We detect alpha particle from the decay of radon daughters. Radon concentration was estimated using energy region of alpha from $^{214}$Po decay in the spectrum and collection efficiency that has dew point dependence in the setup.   
The radon detector, filter housing and all devices were vacuumed to less than 10 Pa. After pure air guaranteed low radon level was filled in the setup by 0.01 MPaG. 

After the preparation, measurement of the $^{222}$Rn emanation was started.
The air inside the detector and sample vessel was circulated by air circulation pump of about 0.8 L/min. The measurement was continued for about 20 days. The right plot on Figure~\ref{fig:ema} demonstrates the time profile for $^{222}$Rn emanation measurement. The emanation rate was estimated from time variation of radon rate with a remained background in radon detector by following equation:
\begin{equation}
R=A\cdot(1-e^{-t/\tau})+B\cdot e^{-t/\tau}, 
\label{eq:fitting}
\end{equation}  
where $R$ is the observed radon rate. $A$ denotes the radon emanation rate from the material. Additionally, $B$ denotes the remaining radon rate in the detector at the beginning of the measurement ($\tau = 3.82/ln2 = 5.52$ days is the lifetime of $^{222}$Rn), and $t$ denotes the time since the start of the measurement. The time profile for radon is fitted by applying Equation~\ref{eq:fitting}.  The radon emanation $A$ of sample 1b filter is estimated as 2.1$\pm$0.1 mBq. Background, which is emanated from the radon detector and setup, is independently measured without a sample less than 0.1 mBq. The decrease from the activity of $^{226}$Ra in HPGe measurement is due to the effect of radon diffusion in the MS.        

\begin{table*}[htbp]
\label{table:devMS4A}
\begin{center}
\caption{The radioactivities of the developed MS4A measured by applying an HPGe detector.}
\begin{tabular}{lcc}
    \hline \hline
    sample name&$^{226}$Ra[mBq/kg]&$^{232}$Th[mBq/kg]\\
    \hline  
   sample 1&22.6$\pm$7.9&91.1$\pm$8.9\\
   sample 1b (solidifying) & 57.0$\pm$14.0&198.4$\pm$16.5\\
   Sample 2&22.8$\pm$9.2&92.4$\pm$10.4\\
    \hline \hline
\end{tabular}
\end{center}
\end{table*}

\section{Conclusion}
\label{sec:conc}
This study developed a very low radioactive MS that can be applied for dark matter search experiments. The MS manufactured is 4A type. The concentration of $^{226}$Ra and $^{232}$Th in powder type is about 57 and 198 mBq/kg measured with HPGe detector, respectively. The reduction of the radioactivity from the commercial MS is about 99 and 97 percent for Ra and Th, respectively. The estimated radon emanation rate from MS filter is about 2.1 mBq measured with a radon detector. This study demonstrated that low RI MS can be made by applying material-screening techniques.

Further work is needed to decrease further the radioactivity of the MS. 
$^{226}$Ra and $^{232}$Th still remained in the developed MS. As shown in Table~\ref{table:selRI}, the silica component is main source of RI. We will search for another silica component with low RI. In addition, the 5A type MS with $\sim$ 5~\AA~pore size, which is expected for the removal of radon from gas, shown in \cite{CYG}, can be produced by ion-exchange of 4A with Ca compound. Consequently, it is important to obtain or synthesize Ca compound of very low activity.

\section*{Acknowledgments}
We gratefully acknowledge the cooperation of Organo Corporation and the cooperation of Kamioka Mining and Smelting Company. 
This work was supported by the Japanese Ministry of Education, Culture, Sports, Science and Technology, the joint research program of the Institute for Cosmic Ray Research (ICRR), Grant-in-Aid for Scientific Research, JSPS KAKENHI Grant Number, 19GS0204, 19H05806, 19K03893 and XMASS collaboration.

%\end{linenumbers}

\begin{thebibliography}{00}
\bibitem{Aprile} E. Aprile and T. Doke, \emph{Liquid xenon detectors for particle physics and astrophysics}, \emph{Rev.Mod.Phys.}, {\bf 82} (2010) 2053-2097.
\bibitem{MEG} S. Mihara et al., \emph{Development of a method for liquid xenon purification using a cryogenic centrifugal pump}, \emph{Cryogenics} {\bf 46} (2006) 688-693.
\bibitem{CYG} A.C. Ezeribe et al., \emph{Demonstration of radon removal from SF$_{6}$ using molecular sieves}, 2017 \emph{JINST} {\bf 12} P09025.
\bibitem{Union} Website for Union Showa K.K., http://www.uskk.co.jp/
\bibitem{RnDet} K.Hosokawa et al., \emph{Development of a high-sensitivity 80 L radon detector for purified gases}, \emph{Prog. Theor. Exp. Phys.} (2015) 033H01.
\end{thebibliography}
\end{document}